\documentclass[12pt]{article}
\usepackage{amssymb}

\def\la{\lambda}
\def\al{\alpha}
\def\de{\delta}

\def\eps{\epsilon}
\def\gam{\gamma}
\newcommand{\exval}[1]{\mbox{$\langle \, {#1}\, \rangle$}}
\newcommand{\bra}[1]{\mbox{$\langle \, {#1}\, |$}}
\newcommand{\ket}[1]{\mbox{$| \, {#1}\, \rangle$}}
\newcommand{\bel}[1]{\begin{equation}\label{#1}}
\newcommand{\be}{\begin{equation}}
\newcommand{\ee}{\end{equation}}
\newcommand{\ba}{\begin{array}}
\newcommand{\ea}{\end{array}}
\newcommand{\bea}{\begin{eqnarray}}
\newcommand{\eea}{\end{eqnarray}}

\begin{document}
\begin{titlepage}             %
\thispagestyle{empty}         %
\begin{center}                %
\vspace*{1cm}                 %
{\large                       
{\bf A sufficient criterion for integrability of stochastic many-body dynamics 
and quantum spin chains\\
}
}\vspace{3cm}                 %
{\large {\sc                  
V. Popkov$^{1,4}$,  M.E. Fouladvand$^{2,3}$ and G.M. Sch\"utz$^{1}$
}}\\[8mm]                     %
{\em
$^1$Institut f\"ur Festk\"orperforschung, Forschungszentrum J\"ulich,
52425 J\"ulich, Germany\\
$^2$Department of physics, Zanjan university, P.O. box 313, Zanjan, Iran\\
$^3$Institute for Studies in Theoretical Physics \& Mathematics(IPM), 
P.O.Box 19395-5531, Tehran,Iran\\
$^4$ Institute for Low Temperature Physics, 310164 Kharkov, Ukraine
}
\vspace{5cm}\\                %
\begin{minipage}{12cm}{\small\sl\rm 
We propose a dynamical matrix product ansatz describing the stochastic 
dynamics of two species of particles with excluded-volume interaction and the
quantum mechanics of the associated quantum spin chains respectively. The
time-dependent algebra which is obtained from the action of the Markov
generator of the exclusion process (or quantum Hamiltonian of the spin chain
respectively) is given in terms of a set of quadratic relations. Analyzing the
permutation consistency of the induced cubic relations we obtain sufficient 
conditions on the hopping rates (i.e., the quantum mechanical interaction
constants) which allow us to identify integrable models. From the dynamical
algebra we construct the quadratic algebra of Zamolodchikov type,
associativity of which is a
 Yang Baxter equation. We also obtain directly from 
the dynamical matrix product ansatz the Bethe ansatz equations for the
spectra of these models.  }
\end{minipage}
\end{center}
\end{titlepage}           %

\section{Introduction and summary of results}
\label{Intro}

The notion of integrability in classical stochastic interacting particle
systems derives from the mapping to quantum spin systems being associated with
a  solution of the Yang-Baxter equation (YBE) \cite{Baxt82,YBE}. Classical 
particle  occupation numbers are interpreted as quantum mechanical spin
degrees of freedom. The generator of the infinitesimal Markovian
time  evolution thus becomes the quantum Hamiltonian of related quantum spin
chain \footnote{The time evolution operators of nonequilibrium systems
correspond to non-Hermitian variants of quantum spin systems, but in the cases
we have in mind this does not affect the integrability.}. In its essence
integrability refers to the existence of an infinite set of conservation laws
which commute with the quantum Hamiltonian and which therefore govern the time
evolution of the stochastic process. For integrable processes the Bethe ansatz
and related methods complement probabilistic approaches \cite{Ligg99} and
allow for the derivation of critical exponents, density profiles and
correlations, shock dynamics and other quantities of interest in the
investigation of driven diffusive systems, reaction-diffusion processes and
other systems both in and and far from equilibrium \cite{Schu00}. Thus
integrability has emerged as a powerful tool in the study of Markov processes
involving one or more species of particles which move and interact a
one-dimensional lattice. Examples include the asymmetric exclusion process,
spin-relaxation processes and reaction-diffusion systems. 

A major difficulty in making use of integrability is posed by the problem of
determining whether a given stochastic dynamics actually does correspond to an
integrable system since the dynamics only determine the quantum Hamiltonian,
but does not directly lead to the Yang-Baxter equation. The same problem
occurs in the context of quantum spin chains alone, i.e., without reference to
stochastic dynamics. Given the Hamiltonian densities as specified by the local
interaction constants one would like to know whether the system is integrable
or not. Even though in special cases an answer can be given by way of
straightforward coordinate Bethe ansatz \cite{Beth31,Yang67},
Baxterization \cite{Baxt} or by an integrability criterion due to Reshetikhin
\cite{Resh}, a generic method for the construction of the Yang-Baxter equation
directly from the quantum Hamiltonian is desirable. It is the purpose of this
paper to provide such a method for stochastic interacting particle systems and
their associated quantum spin chain Hamiltonians.

To this end we extend the dynamical matrix product ansatz (DMPA) 
\cite{Stin95a,Stin95b}, originally developed for the single-species
exclusion process and the associated spin-1/2 Heisenberg quantum chain,
to two-species exclusion process with nearest-neighbor interaction and their
associated spin 1 quantum chains. This is an approach where the action
of the time evolution operator on an arbitrary state is rephrased in terms
of a dynamical algebra of time-dependent operators with
quadratic relations (see below). In this way we are able to answer two
 questions. The first is: Since the DMPA can be constructed for
any stochastic process with nearest-neighbor interaction, what is special
about the resulting algebra in integrable cases? 
We find that the special property is  associativity
which in its turn 
requires that cubic relations of the operators satisfy the YBE. The
second question is: What choice of interaction parameters leads to an integrable case?
Since our method is constructive and yields the YBE without a prior assumption
on parameters,
we obtain relations
between the parameters which provide a {\em sufficient} criterion for
integrability directly from the quantum Hamiltonian (i.e., without explicit
reference to the infinite set of conservation laws and to an underlying 
transfer-matrix). In this way we recover
not only a known family of integrable two-species exclusion processes which,
using the Reshetikhin criterion, was believed to be the only one \cite{ADR},
but also a new family which had been studied earlier \cite{xxx}, but had not
been known to be integrable.

This paper addresses rather different communities, viz. physicists and
mathematicians working on interacting particle systems on the one hand and
on integrable models on the other hand. For the first group some of the
remarks on quantum spin chains and integrable quantum field theory which
appear in the paper may seem obscure and perplexing. Therefore we have tried 
to give a self-contained presentation of our results such that the paper can 
be followed even if all these allusions are being ignored. For a detailed
and more pedagogical review we refer to \cite{Schu00}. Probabilists may also find
the mathematically rigorous application of related quantum techniques in 
\cite{Sudb95,Beli01} useful in this respect. We hope that for the second
group of readers the non-standard (and non-rigorous) application of quantum
integrability to classical stochastic dynamics pursued here will prove
inspiring and not distracting. Complementary to the review \cite{Schu00} we
refer to \cite{Ligg99,Spoh91,Ligg85,Priv97} for the significance of stochastic
interacting particle systems in a wider mathematical and physical  perspective
and to \cite{Schm95} specifically for driven lattice gases of the type studied
here.  

The paper is organized as follows. In Sec. II we first follow standard
procedure \cite{Schu00} and introduce the quantum spin chain representation of
the stochastic many-body dynamics. Then  (Sec. III) we extend the DMPA
developed in \cite{Stin95a,Stin95b} for the single-species exclusion process to
the two-species case. The special case corresponding to symmetric hopping has
been discussed recently in \cite{Stin01}. In Sec. IV we derive the YBE from 
the dynamical algebra and obtain the conditions on the parameters which
guarantee that the YBE is satisfied. In Sec. V we derive the nested Bethe
ansatz equations for the spectrum of the quantum Hamiltonian. 
The idea of proposing matrix states as {\it special} eigenstates of
periodic quantum systems was first employed in \cite{Affleck,KSZ91}.
 

\section{Quantum Hamiltonian formalism}
\label{QHF}

The method of defining stochastic dynamics in terms of an imaginary-time
Schr\"odinger equation with a ``quantum Hamiltonian'' as generator of the
Markovian stochastic dynamics has been reviewed in detail in \cite{Schu00}.
In order to be self-contained we present here a summary of what is
required below. 

Consider a configuration 
$\eta$ of particles on a lattice of $L$ sites. Here $\eta = \{\eta(1), 
\eta(2), \dots , \eta(L)\}$ where $\eta(x)$ denotes the state of the system
at site $x$ in terms of an occupation number for each permissible
particle species. For definiteness we consider here only two-species
exclusion processes where each lattice site is either occupied
with a particle of species $A$ (local occupation number $n_x^A=1$) or species
$B$ ($n_x^B=1$) or vacant respectively ($n_x^A=n_x^B=0$).
We may therefore define the ``spin'' variable $\eta(x) = n_x^A - n_x^B \in
\{1,0,-1\}$ as a unique variable specifying site $x$. Physically one may
interprete $\eta(x)$ e.g. as the charge of the particle occupying site $x$.

In the course of time random events take place which change a configuration
$\eta$ of the system with rate $w_{\eta\to\eta'}$.
We define a stochastic process in terms of a master equation 
\bel{2-1}
\frac{d}{dt} P(\eta;t) = \sum_{\stackrel{\eta'\in X}{\eta'\neq\eta}}
\left[w_{\eta'\to\eta}P(\eta';t) -
w_{\eta\to\eta'}P(\eta;t)\right]
\ee
for the probability $P(\eta;t)$ of finding the state $\eta$ at time $t$.

The idea of the ``quantum Hamiltonian'' formalism is to
represent each of the possible particle configurations $\eta$ by a vector
$\ket{\eta}$ which together with the transposed vectors  $\bra{\eta}$ form a
basis of the vector space $X=({\mathbb{C}}^2)^{\otimes L}$ and its dual
respectively with scalar product $\bra{\eta}\,\eta'\,\rangle=
\delta_{\eta,\eta'}$. We represent an empty site by the symbol $0$
and occupied sites by $A,B$ and choose as local basis vectors
\bel{2-2}
\ket{0} = \left(\ba{c} 1 \\ 0 \\ 0 \ea \right),\quad
\ket{A} = \left(\ba{c} 0 \\ 1 \\ 0 \ea \right),\quad
\ket{B} = \left(\ba{c} 0 \\ 0 \\ 1 \ea \right).
\ee
A state $\eta$ of the entire system is then represented by a tensor state
$\ket{\eta} =  \ket{\eta_1}\otimes\dots\otimes\dots\ket{\eta_L}$.
Therefore the probability distribution is given
by a state  vector
\bel{2-3}
| \, P(t)\, \rangle = \sum_{\eta \in X} P(\eta;t) \ket{\eta}.
\ee
In this formalism one rewrites the master equation in the form of
a Schr\"odinger equation in imaginary time,
\bel{2-4}
\frac{d}{dt} P(\eta;t) = - \langle \, \eta \, |
H | \, P(t) \, \rangle ,
\ee
where the off-diagonal matrix elements of $H$ are the (negative) transition 
rates $w_{(\eta'\to(\eta}$ between states and the diagonal entries are the
inverse of the exponentially distributed life times of the states, i.e.,
the sum of all outgoig transition rates $w_{\eta\to\eta'}$ from state $\eta$.
In quantum mechanical interpretation $\eta(x)$ may be regarded as the
$z$-component of the spin of an atom in a spin 1 state.

The master equation is linear in time and therefore has a formally simple
solution. A state at time $t=t_0 + \tau$ is given in terms of an initial
state at time $t_0$ by
\bel{2-5}
| \, P(t_0+\tau) \, \rangle = \mbox{e}^{-H\tau } 
| \, P(t_0) \, \rangle.
\ee 
The expectation value $\rho_k(t)=\bra{s} n_k^Z \ket{P(t)}$, $Z=A,B$, for the
$Z$-species density at  site $x$ is given by the projection operator $n_k^Z$
which has value 1 if  there is a particle of type $Z$ at site $k$ and 0
otherwise. The constant vector  $\bra{s} = \sum_{\eta \in X} \bra{\eta}$ performs the
average over all possible final states of the stochastic time evolution.
The real part of the spectrum of $H$ is the set of all inverse relaxation 
times of the process. A non-vanishing imaginary part signals the presence
of currents characterizing an non-equilibrium system. An equilibrium system
satisfying detailed balance has a purely real relaxation spectrum
and can always be mapped to a quantum
system with real symmetric Hamiltonian $H$.

Here we consider diffusive systems with only hopping processes between
neighboring sites. We assign hopping rates as follows
\be
\label{2-6}
\ba{lllll}
A0 & \to & 0A & \mbox{with rate} & g_{A0} \\
0A & \to & A0 &                  & g_{0A} \\
B0 & \to & 0B &                  & g_{B0} \\
0B & \to & B0 &                  & g_{0B} \\
AB & \to & BA &                  & g_{AB} \\
BA & \to & AB &                  & g_{BA}.
\ea
\ee
The transition matrix for the process defined on two lattice sites takes
the form
\be
\label{2-7}
h = \left( \ba{ccccccccc}
0 &   0   &   0   &   0   & 0 &   0   &   0   &   0   & 0 \\
0 & g_{0A}&   0   &-g_{A0}& 0 &   0   &   0   &   0   & 0 \\
0 &   0   & g_{0B}&   0   & 0 &   0   &-g_{B0}&   0   & 0 \\
0 &-g_{0A}&   0   & g_{A0}& 0 &   0   &   0   &   0   & 0 \\
0 &   0   &   0   &   0   & 0 &   0   &   0   &   0   & 0 \\
0 &   0   &   0   &   0   & 0 & g_{AB}&   0   &-g_{BA}& 0 \\
0 &   0   &-g_{0B}&   0   & 0 &   0   & g_{B0}&   0   & 0 \\
0 &   0   &   0   &   0   & 0 &-g_{AB}&   0   & g_{BA}& 0 \\
0 &   0   &   0   &   0   & 0 &   0   &   0   &   0   & 0 \ea \right).
\ee
In a periodic system with $L$ sites one therefore has
\bel{2-8}
H = \sum_{i=1}^L h_{i}
\ee
where $h_{i} = {\bf 1} \otimes \dots  \otimes h \otimes\dots  \otimes{\bf 1}$
is the hopping matrix (\ref{2-7}) acting on sites $i,i+1$ and ${\bf 1}$
is the $3 \times 3$ unit matrix. For systems with open boundaries
where particles are exchanged with external reservoirs the hopping
matrix $h_L$ is replaced by suitably chosen boundary matrices $b_{1},b_L$.

\section{Dynamical Matrix Product Ansatz}
\label{DMPA}

In order to introduce the dynamical matrix product ansatz for a
three-state system we define matrix valued  vectors for a single site
\bea
\label{3-1}
\ket{\bf A} & = & \left(\ba{c} E \\ D^A \\ D^B \ea \right) \\
\label{3-2}
\ket{\bf X} & = & \left(\ba{c} X^0 \\ X^A \\ X^B \ea \right) 
\eea
with time-dependent matrices $E$, $D^A$, $D^B$ and the  time-dependent
auxiliary matrices $X^{0,A,B}$. By taking a $L$-fold tensor product one
obtains a matrix product state (MPS) 
\bel{3-3}
\ket{{\bf P}(t)} = \left(\ba{c} E \\ D^A \\ D^B \ea \right)^{\otimes L-1}
\otimes \left(\ba{c} EQ \\ D^A Q \\ D^B Q \ea  \right).
\ee

These matrices are chosen such that the MPS satisfies the master equation
(\ref{2-4}) for the hopping process defined by (\ref{2-6}).
Notice that there is  the multiplication by some additional matrix $Q$ of the $L$-th
term in (\ref{3-3}). Choosing $Q$ to be time-independent,
$\dot Q=0$, and 
 following
\cite{Stin95a} this leads to an infinite-dimensional algebra of the matrices
introduced above and their time derivatives which has quadratic relations
given by 
\bel{3-4}
\left( \frac{1}{2}\frac{d}{dt} + h \right) \ket{\bf A} \otimes \ket{\bf A}
= \ket{\bf X}\otimes \ket{\bf A} -  \ket{\bf A} \otimes \ket{\bf X}.
\ee
Special care has to be given to the action of terms
$h_L, h_{L-1}$ of the Hamiltonian  (\ref{2-8}) on $P(t)$, due to matrix
$Q$ involved in (\ref{3-3}). It can be shown, however, \cite{elsewhere}
that with $Q$ satisfying
\bel{Q_prop1}
[Q,E^{-1}D^Z]=[Q,D^ZE^{-1}]=0; \ \ \ Z=A,B.
\ee
(we assume further invertibility of $E$), no extra relations arise in 
addition to (\ref{3-4}). 
Note that in general we assume that  
$Q$  commutes neither with $D$ nor with $E$ separately, 
$[Q,E]\neq 0; \ \  [Q,D]\neq 0$. 

Eqs. (\ref{3-4}) contain nine relations, 
namely:
\bea
\label{algebra2a}
\frac{1}{2}\left(\dot{E}E + E\dot{E}\right) & = & X^0E - EX^0 \\
\label{algebra2b}
\frac{1}{2}\left(\dot{E}D^A + E\dot{D}^A\right) + g_{0A} ED^A - g_{A0} D^A E 
                                            & = & X^0D^A - EX^A \\
\label{algebra2c}
\frac{1}{2}\left(\dot{E}D^B + E\dot{D}^B\right) + g_{0B} ED^B - g_{B0} D^B E 
                                            & = & X^0D^B - EX^B \\
\label{algebra2d}
\frac{1}{2}\left(\dot{D}^AE + D^A\dot{E}\right) - g_{0A} ED^A + g_{A0} D^A E 
                                            & = & X^A E - D^A X^0 \\
\label{algebra2e}
\frac{1}{2}\left(\dot{D}^AD^A + D^A\dot{D}^A\right) & = & X^AD^A - D^AX^A \\
\label{algebra2f}
\frac{1}{2}\left(\dot{D}^AD^B + D^A\dot{D}^B\right) 
                   + g_{AB} D^AD^B - g_{BA} D^B D^A & = & X^AD^B - D^AX^B \\
\label{algebra2g}
\frac{1}{2}\left(\dot{D}^BE + D^B\dot{E}\right) - g_{0B} ED^B + g_{B0} D^B E 
                                            & = & X^BE - D^BX^0 \\
\label{algebra2h}
\frac{1}{2}\left(\dot{D}^BD^A + D^B\dot{D}^A\right) 
                   - g_{AB} D^AD^B - g_{BA} D^B D^A & = & X^BD^A - D^BX^A \\
\label{algebra2i}
\frac{1}{2}\left(\dot{D}^BD^B + D^B\dot{D}^B\right) & = & X^BD^B - D^BX^B
\eea

If one finds matrices satisfying these relations one can calculate
expectation values and probabilities $P(\eta,t)$ by taking a trace 
in the space on which the matrices act, i.e.
\be
\ket{P(t)} = \mbox{Tr } \ket{{\bf P}(t)}/Z_L
\ee
Here $Z_L = \mbox{Tr } C^L Q$ with $C=E+D^A+D^B$. This is the sum of all
unnormalized configurational probabilities and hence yields the correct
normalization factor. {}For a given configuration $\eta$ the probability
$P(\eta,t)$ is therefore obtained by taking the trace over a
suitably chosen normalized product of $L$ matrices $D^A,D^B,E$. One represents
an occupied (vacant) site by a time-dependent matrix $D^Z$ ($E$) in a string 
$DDDEDEE\dots$ of $L$ such matrices. For
illustration consider the probability $\exval{n^Z_xn^{Z'}_y(t)}$ of finding
two particles of species $Z$, $Z'$ at sites $x,y$ in an otherwise empty 
system. One has 
\bel{exval}
\exval{n^Z_x(t)n^{Z'}_y(t)} = \mbox{Tr } (E^{x-1} D^Z E^{y-x-1} D^{Z'} 
E^{L-y} Q)/Z_L.  
\ee
The initial state
of the system is encoded in the initial values at $t=0$ of the matrices.
 This time-dependent algebra generalizes the stationary 
two-species case studied in detail in \cite{Arnd98} and the time-dependent
single-species algebra introduced in \cite{Stin95a}. 

The algebra can be exploited either by studying explicit matrix 
representations or on a purely algebraic level \cite{Schu98}. Here we choose
the second approach which has two objectives: (i) the elimination of the
auxiliary operators in such a manner that information necessary to
calculate the spectrum of $H$ as well as the configurational probabilities is
retained, (ii) the elimination of the time-dependence from the algebraic
relations. Our analysis consists of several steps, each of which leads
to a reduction of the full algebra (\ref{algebra2a}) - (\ref{algebra2i})
to a smaller algebra with fewer generators and fewer relations.\\[4mm]
{\it Step 1:}\\[4mm]
Choosing
\bel{ansatz1}
\dot{E} = 0
\ee
one can without loss of generality satisfy (\ref{algebra2a}) with the choice
\bel{X0}
X^0 = 0
\ee
One has the remaining eight relations 
\bea
\label{algebra3b}
\frac{1}{2}\left(E\dot{D}^A\right) + g_{0A} ED^A - g_{A0} D^A E 
                                            & = & - EX^A \\
\label{algebra3c}
\frac{1}{2}\left(E\dot{D}^B\right) + g_{0B} ED^B - g_{B0} D^B E 
                                            & = & - EX^B \\
\label{algebra3d}
\frac{1}{2}\left(\dot{D}^AE \right) - g_{0A} ED^A + g_{A0} D^A E 
                                            & = & X^A E \\
\label{algebra3e}
\frac{1}{2}\left(\dot{D^A}D^A + D^A\dot{D^A}\right) & = & X^AD^A - D^AX^A \\
\label{algebra3f}
\frac{1}{2}\left(\dot{D}^AD^B + D^A\dot{D}^B\right) 
                   + g_{AB} D^AD^B - g_{BA} D^B D^A & = & X^AD^B - D^AX^B \\
\label{algebra3g}
\frac{1}{2}\left(\dot{D}^BE\right) - g_{0B} ED^B + g_{B0} D^B E 
                                            & = & X^BE  \\
\label{algebra3h}
\frac{1}{2}\left(\dot{D}^BD^A + D^B\dot{D}^A\right) 
                   - g_{AB} D^AD^B - g_{BA} D^B D^A & = & X^BD^A - D^BX^A \\
\label{algebra3i}
\frac{1}{2}\left(\dot{D^B}D^B + D^B\dot{D^B}\right) & = & X^BD^B - D^BX^B
\eea
A similar constraint was used in Ref. \cite{Sasa} for the study of the
single-species case, leaving three instead of four relations. In 
\cite{Stin01} the choice $\dot{C}=0$ is made for the two-species case.
In this case the sum of the auxiliary matrices is set to zero which
has been shown to involve no loss of generality \cite{Sant97}.\\[4mm]
{\it Step 2:}\\[4mm]
Adapting the strategy of Ref. \cite{Stin95a} we further
assume that $E$ is invertible. This allows
us to express the remaining auxiliary matrices $X_{A,B}$ in terms of the
physical matrices $E,D^A,D^B$. Multiplying eqs. (\ref{algebra3b}),
(\ref{algebra3c}),(\ref{algebra3d}),(\ref{algebra3g}) from the right and
left resp. with $E^{-1}$ we find:
\bea
\label{XA}
2 X^A & = & g_{A0}(D^A+E^{-1}D^A E) - g_{0A}(D^A+ED^A E^{-1}) \\
\label{XB}
2 X^B & = & g_{B0}(D^B+E^{-1}D^B E) - g_{0B}(D^B+ED^B E^{-1}) 
\eea

Now there are six relations left. Two involve the time-derivatives
of $D^A$ and $D^B$ respectively:
\bea
\label{algebra4a}
\dot{D}^A & = & - g_{A0}(D^A-E^{-1}D^A E) - g_{0A}(D^A-ED^A E^{-1}) \\
\label{algebra4b}
\dot{D}^B & = & - g_{B0}(D^B-E^{-1}D^B E) - g_{0B}(D^B-ED^B E^{-1}) 
\eea
Using also this yields four other equations:
\bea
\label{algebra4c}
g_{A0} D^A E^{-1}D^A E + g_{0A} ED^A E^{-1}D^A & = &
                                    (g_{A0}+ g_{0A}) (D^A)^2 \\
\label{algebra4d}
g_{B0} D^B E^{-1}D^B E + g_{0B} ED^B E^{-1}D^B & = &
                                    (g_{B0}+ g_{0B}) (D^B)^2 \\
\label{algebra4e}
g_{0A} ED^A E^{-1}D^B + g_{B0} D^A E^{-1}D^B E & = &
            (g_{A0}+ g_{0B}-g_{AB}) D^AD^B + g_{BA} D^BD^A  \\
\label{algebra4f}
g_{0B} ED^B E^{-1}D^A + g_{A0} D^B E^{-1}D^A E & = &
            (g_{B0}+ g_{0A}-g_{BA}) D^BD^A + g_{AB} D^AD^B .
\eea
Thus the originally quadratic problem of the time-evolution 
with nine relations for three physical
operators and three auxiliary operators has been converted into the quartic
problem (\ref{algebra4a}) - (\ref{algebra4f}) with six relations for the
physical operators alone which have to be satisfied at all times. Notice
that the relations (\ref{algebra4a}), (\ref{algebra4b}) are linear in the
$D$-operators while the relations (\ref{algebra4c}) - (\ref{algebra4f}) are
bilinear.\\[4mm]
{\it Step 3:}\\[4mm]
The linear relations (\ref{algebra4a}), (\ref{algebra4b}) are sufficient
to describe the one-particle sector, i.e. the set of configurations with
only one particle on the lattice. In order to eliminate the time-dependence
we formally define ``Fourier'' components 
\bel{FAB}
{\cal D}^Z_p(t)  = \sum_{k=-\infty}^{\infty}
\mbox{e}^{ipk} D^Z_k(t)
\ee

\bel{DkAB} 
D^A_k(t)= \alpha^k  E^{k-1}D^A(t) E^{-k};\ \ \ 
D^B_k(t)= \beta^k  E^{k-1}D^A(t) E^{-k}
\ee
which have the property
\bea
\label{TA}
E^{-1} {\cal D}^{A}_p(t) E & = & \alpha \mbox{e}^{ip} {\cal D}^{A}_p(t) \\
\label{TA1}
E^{-1} {\cal D}^{B}_p(t) E & = & \beta  \mbox{e}^{ip} {\cal D}^{B}_p(t) 
\eea
Additionally we require that the matrix $Q$ from (\ref{3-3}),(\ref{exval})
obeys
\bel{Q_prop}
[Q,{D}_k^Z(t)]=0
\ee
which contains (\ref{Q_prop1}) as a special case and yields
\bel{QDp}
[Q,{\cal D}^Z_p (t)]=0 
\ee 
Conversely one has
\bel{INVFAB}
D_k^{A,B}(t) = \frac{1}{2\pi} \int_{-\pi}^{\pi} 
\mbox{e}^{-ipk} {\cal D}^{A,B}_p(t)
 \mbox{d}p.
\ee

The Fourier ansatz turns the time-dependent relations (\ref{algebra4a}), 
(\ref{algebra4b}) into two ordinary first-order differential equations
\be
\dot{\cal D}^{A,B}_p(t) = - \epsilon^{A,B}_p {\cal D}^{A,B}_p(t)
\ee
with the ``dispersion relations''
\bea
\label{dispersionA}
\epsilon^{A}_p & = & g_{0A} \alpha^{-1} \mbox{e}^{-ip} + g_{A0} \alpha
\mbox{e}^{ip} -  g_{0A} - g_{A0} \\
\label{dispersionB}
\epsilon^{B}_p & = & g_{0B} \beta^{-1} \mbox{e}^{-ip} + g_{B0} \beta
\mbox{e}^{ip} -  g_{0B} - g_{B0}.
\eea
In this way one can express the time-dependent matrix ${\cal D}^{A,B}_p(t)$
in terms of its initial value as
\bel{solution}
{\cal D}^{A,B}_p(t) = \mbox{e}^{- \epsilon^{A,B}_pt} {\cal D}^{A,B}_p(0).
\ee
In what follows we shall omit the time-argument in the initial matrices
$D^Z_{p}(0)$.

In terms of the Fourier components (\ref{FAB}) the four relations
(\ref{algebra4c}) - (\ref{algebra4f}) turn into double-integral relations
where the time-dependence is shuffled into an exponential. Because of
(\ref{TA}) the quartic relations turn into quadratic relations. Using
(\ref{INVFAB}) with $p_1$ for the $A$ species and $p_2$ for the $B$ species
in the third relation and vice versa in the fourth relation below
one gets
\bea
\label{algebra5a}
0 & = & \int \mbox{d}p_1 \int \mbox{d}p_2 a_{12} {\cal D}^A_{p_1}
{\cal D}^A_{p_2} \mbox{e}^{-(\epsilon^A_{p_1}+\epsilon^A_{p_2})t}\\
\label{algebra5b}
0 & = & \int \mbox{d}p_1 \int \mbox{d}p_2 b_{12} {\cal D}^B_{p_1}
{\cal D}^B_{p_2} \mbox{e}^{-(\epsilon^B_{p_1}+\epsilon^B_{p_2})t}\\
\label{algebra5c}
0 & = & \int \mbox{d}p_1 \int \mbox{d}p_2 [c_{12} {\cal D}^A_{p_1}
{\cal D}^B_{p_2} - g_{BA} \alpha^{-1} \mbox{e}^{-ip_1} 
{\cal D}^B_{p_2}{\cal D}^A_{p_1}]
\mbox{e}^{-(\epsilon^A_{p_1}+\epsilon^B_{p_2})t}\\
\label{algebra5d}
0 & = & \int \mbox{d}p_1 \int \mbox{d}p_2[d_{12} {\cal D}^B_{p_1}
{\cal D}^A_{p_2} - g_{AB} \beta^{-1}\mbox{e}^{-ip_1} 
{\cal D}^A_{p_2}{\cal D}^B_{p_1}]
\mbox{e}^{-(\epsilon^B_{p_1}+\epsilon^A_{p_2})t}
\eea
with the functions
\bea
\label{a12}
a_{12} \equiv a(p_1,p_2) & = & g_{0A} \alpha^{-2}\mbox{e}^{-ip_1-ip_2}
+ g_{A0} - (g_{0A} + g_{A0})\alpha^{-1}\mbox{e}^{-ip_2}\\
\label{b12}
b_{12} \equiv b(p_1,p_2) & = & g_{0B} \beta^{-2}\mbox{e}^{-ip_1-ip_2}
+ g_{B0} - (g_{0B} + g_{B0})\beta^{-1}\mbox{e}^{-ip_2}\\
\label{c12}
c_{12} \equiv c(p_1,p_2) & = & g_{0A} \alpha^{-1}\beta^{-1}\mbox{e}^{-ip_1-ip_2}
+ g_{B0} - (g_{0B} + g_{A0}-g_{AB})\beta^{-1}\mbox{e}^{-ip_2}\\
\label{d12}
d_{12} \equiv d(p_1,p_2) & = & g_{0B} \alpha^{-1}\beta^{-1}\mbox{e}^{-ip_1-ip_2}
+ g_{A0} - (g_{0A} + g_{B0}-g_{BA})\alpha^{-1}\mbox{e}^{-ip_2}
\eea
The four integral equations (\ref{algebra5a}) - (\ref{algebra5d}), obtained
from  (\ref{ansatz1}) together with the assumption of existence of $E^{-1}$ 
and with  the choice (\ref{X0}), form the basis of the subsequent analysis.
As an intermediate summary we remark that at this point the only relations
(out of originally nine) that remain 
are four bilinear relations (\ref{algebra5a}) - (\ref{algebra5d}).
Correspondingly, all expectation values can be calculated from the
initial matrices using (\ref{exval}), (\ref{INVFAB}), (\ref{solution}).
E.g. one has (see (\ref{exval}),(\ref{DkAB}),(\ref{INVFAB}))
\be
\exval{n^Z_x(t)n^{Z'}_y(t)} = 
\mbox{Tr } (D_x^Z D_y^{Z'}E^L Q)/Z_L=
\int \int dp_1 dp_2 
\mbox{e}^{-(\epsilon^Z_{p_1}+\epsilon^{Z'}_{p_2})t} \mbox{e}^{-ip_1x-ip_2y}
\mbox{Tr } ({\cal D}^Z_{p_1}{\cal D}^{Z'}_{p_2}E^L Q)/Z_L.
\ee
The only unknown quantities are time-independent matrix product elements of 
the form $\mbox{Tr } (D^Z_{p_1}(0) D^{Z'}_{p_2}(0)\dots  E^{L} Q)/Z_L$, to be
discussed below. \\[4mm]
{\it Step 4:}\\[4mm]
Before proceeding further a distinction between two different cases
must be made. Since we assume  that no particle species is completely
immobile (i.e. the possibilities $g_{A0}=g_{0A}=0$ or $g_{B0}=g_{0B}=0$
are excluded) the functions $a_{12}$ and $b_{12}$ do not vanish identically.
Indeed in the generic case, in the following referred to as case I, none
of the four integrands (\ref{algebra5a}) - (\ref{algebra5d}) vanish
identically. Only for the special case
$0=g_{0A}=g_{B0}=g_{BA}=g_{AB}-g_{A0}-g_{0B}$ corresponding to
$c_{12}=g_{BA}=0$ (or the equivalent case obtained by interchanging $A$ and
$B$ species) the integrand in (\ref{algebra5c}) (or (\ref{algebra5d}) resp.) is
zero. This is case II, to be treated separately.
\\[4mm]
{\it Case I:}\\[4mm]
The relations (\ref{algebra5a}) - (\ref{algebra5d}) may be reformulated
by splitting the integral into two parts as
\bel{split}
 \int_{-\pi}^{\pi} \int_{-\pi}^{\pi} F_{p_1,p_2}  \mbox{d}p_1 \mbox{d}p_2=
 \int_{-\pi}^{\pi} \int_{-\pi}^{p_1} \ldots +
\int_{-\pi}^{\pi} \int_{p_1}^{\pi} \ldots= I_{p2<p1}+ I_{p2>p1}  
\ee
By changing the order of integration and interchanging $p_1 \leftrightarrow p_2$
in the last term in (\ref{split}) we obtain
 \bel{split2}
I_{p2>p1}= \int_{-\pi}^{\pi} \int_{-\pi}^{p_2}  F_{p_1,p_2} \mbox{d}p_2 \mbox{d}p_1 =
\int_{-\pi}^{\pi} \int_{-\pi}^{p_1}  F_{p_2,p_1} \mbox{d}p_1 \mbox{d}p_2
\ee
Using (\ref{split2},\ref{split}) the  relation (\ref{algebra5a}) thus becomes

\be
\int_{- \pi}^{\pi} \mbox{d}p_1 \int_{- \pi}^{p_1} \mbox{d}p_2 [a_{12} {\cal D}^A_{p_1}
{\cal D}^A_{p_2}+ a_{21} {\cal D}^A_{p_2}
{\cal D}^A_{p_1}]\mbox{e}^{-(\epsilon^A_{p_1}+\epsilon^A_{p_2})t} = 0.
\ee
In order to satisfy this for all times $t$
we require the integrand inside the brackets to vanish. This is a sufficient
condition for satisfying also the original equations (\ref{algebra5a}).
A similar condition is obtained for (\ref{algebra5b}). Relations
(\ref{algebra5c}), (\ref{algebra5d}) can also be satisfied in this manner,
but one has to require
\be
\epsilon^A_{p_1}+\epsilon^B_{p_2} =\epsilon^A_{p_2}+\epsilon^B_{p_1}.
\ee
This implies the constraints

\bea
\label{aobo}
\alpha g_{A0} &=& \beta g_{B0} \\
\label{oaob}
\alpha^{-1} g_{0A} &=& \beta^{-1} g_{0B} 
\eea
on the hopping rates. 

In what follows we assume (\ref{aobo}) and  (\ref{oaob}) to
hold. One obtains the following four relations
\bea
\label{algebra6a}
a_{12} {\cal D}^A_{p_1}{\cal D}^A_{p_2} 
& = & - a_{21}{\cal D}^A_{p_2}{\cal D}^A_{p_1} \\
\label{algebra6b}
 b_{12} {\cal D}^B_{p_1} {\cal D}^B_{p_2} 
& = & - b_{21}{\cal D}^B_{p_2}{\cal D}^B_{p_1} \\
\label{algebra6c}
c_{12} {\cal D}^A_{p_1}{\cal D}^B_{p_2} - g_{BA} \alpha^{-1} \mbox{e}^{-ip_1} 
{\cal D}^B_{p_2}{\cal D}^A_{p_1}
& = & - c_{21} {\cal D}^A_{p_2}{\cal D}^B_{p_1} + g_{BA} \alpha^{-1}
\mbox{e}^{-ip_2}  {\cal D}^B_{p_1}{\cal D}^A_{p_2}\\
\label{algebra6d}
d_{12} {\cal D}^B_{p_1} {\cal D}^A_{p_2} - g_{AB} \beta^{-1}\mbox{e}^{-ip_1} 
{\cal D}^A_{p_2}{\cal D}^B_{p_1}
& = & - d_{21} {\cal D}^B_{p_2} {\cal D}^A_{p_1} + g_{AB}
\beta^{-1}\mbox{e}^{-ip_2}  {\cal D}^A_{p_1}{\cal D}^B_{p_2}
\eea
We stress that in these relations the time-dependence drops out.
Therefore these are four static relations on the operators ${\cal D}^Z_{p}(0)$
which together with the two equations (\ref{solution}) and with
(\ref{ansatz1}), (\ref{X0}) form a sufficient (but not necessary) condition
for satisfying the original algebra (\ref{3-4}) on the manifold defined by
(\ref{aobo}),(\ref{oaob}).

These algebraic relations can be written in a compact form by defining the
operator valued two-component vector
\bel{ovec}
{\cal D}_p = \left( \ba{l} {\cal D}^A_p \\{\cal D}^B_p \ea \right)
\ee
and the $4\times 4$ matrix
\bel{sigma}
\Sigma(p_1,p_2) = \left( \ba{cccc} 
\sigma_{AA}^{AA} & 0 & 0 & 0 \\
0 & \sigma_{AB}^{AB} & \sigma_{AB}^{BA} & 0 \\
0 & \sigma_{BA}^{AB} & \sigma_{BA}^{BA} & 0 \\
0 & 0 & 0 & \sigma_{BB}^{BB} 
\ea \right)
\ee
with
\bea
\sigma_{AA}^{AA} & = & -\frac{a_{21}}{a_{12}} \\
\sigma_{AB}^{AB} & = & 
-\frac{\alpha\beta c_{21}d_{12} - g_{AB}g_{BA} \mbox{e}^{-ip_1-ip_2}}
{\alpha\beta c_{12}d_{12} - g_{AB}g_{BA} \mbox{e}^{-2ip_2}} \\
\sigma_{AB}^{BA} & = & 
g_{BA}\beta\frac{\mbox{e}^{-ip_1}d_{12} - \mbox{e}^{-ip_2}d_{21}}
{\alpha\beta c_{12}d_{12} - g_{AB}g_{BA} \mbox{e}^{-2ip_2}} \\
\sigma_{BA}^{AB} & = & 
g_{AB}\alpha\frac{\mbox{e}^{-ip_1}c_{12} - \mbox{e}^{-ip_2}c_{21}}
{\alpha\beta c_{12}d_{12} - g_{AB}g_{BA} \mbox{e}^{-2ip_2}} \\
\sigma_{BA}^{BA} & = & 
-\frac{\alpha\beta d_{21}c_{12} - g_{AB}g_{BA} \mbox{e}^{-ip_1-ip_2}}
{\alpha\beta c_{12}d_{12} - g_{AB}g_{BA} \mbox{e}^{-2ip_2}} \\
\sigma_{AA}^{AA} & = & -\frac{b_{21}}{b_{12}} .
\eea
With these quantities relations (\ref{algebra6a}) - (\ref{algebra6d}) read
\bel{algebra7}
{\cal D}_{p_1} \otimes {\cal D}_{p_2} = \Sigma(p_1,p_2) 
{\cal D}_{p_2} \otimes {\cal D}_{p_1}
\ee
We remark that the Fourier components ${\cal D}^Z_{p}(0)$ satisfy the algebra
of creation operators in a 1+1-dimensional integrable quantum field theory
\cite{Zamolodchikov}.  The matrix $S=\Sigma P$ with the permutation operator $P$ acting
on the two vector spaces may then be regarded as scattering matrix with matrix
elements $S_{YY'}^{ZZ'}$ in row $YY'$ and column $ZZ'$,
\bel{scat}
{\cal D}_{p_1}^Z {\cal D}_{p_2}^{Z'} =
S_{Y Y'}^{Z Z'}  {\cal D}_{p_2}^{Y'} {\cal D}_{p_1}^Y
\ee
 Applying
(\ref{algebra7}) twice shows that $\Sigma$ to satisfy
\bel{unitary}
\Sigma(p_1,p_2) \Sigma(p_2,p_1) = 1
\ee
which is fulfilled without further constraints on the hopping rates. 
This is the analog of the field theoretical unitarity condition.
\\[4mm]
{\it Case II:}\\[4mm]
Case II corresponds to the manifold
\bel{constraint2}
g_{0A}=g_{B0}=g_{BA}=0,\quad g_{AB}=g_{A0}+g_{0B}
\ee
which is the totally asymmetric two-species exclusion process
\bel{constraint2_process}
\ba{lllll}
A0 & \to & 0A & \mbox{with rate} & g_{A0} \\
0B & \to & B0 &                  & g_{0B} \\
AB & \to & BA &                  & g_{A0}+g_{0B}.
\ea
\ee
investigated in \cite{xxx}. By exchanging vacancies with $B$ particles
this process is equivalent to the totally asymmetric process with both
species hopping to the right, but 
\be
\label{alternative}
\ba{lllll}
A0 & \to & 0A & \mbox{with rate} & g_{A0}+g_{0B} \\
B0 & \to & 0B &                  & g_{0B} \\
AB & \to & BA &                  & g_{A0}.
\ea
\ee
On the manifold (\ref{constraint2}) relation (\ref{algebra5c})
is satisfied identically. Without further constraints on the independent
hopping rates $g_{A0}$, $g_{0B}$ and on the free
parameters $\alpha,\beta$ one may satisfy (\ref{algebra5a}),
(\ref{algebra5b}) by splitting the integrals into
two domains as in case I, but by requiring the integrand in 
(\ref{algebra5d}) to vanish {\em without} splitting of the integral.
As in case {\rm I} the resulting algebraic relations take the form
(\ref{scat}) with
\bea
\label{S_caseII_begin}
S_{AA}^{AA} & = & -\frac{a_{21}}{a_{12}}= 
- \frac{\al-e^{-ip_1}} {\al-e^{-ip_2}} \\
S_{AB}^{BA}=S_{BA}^{AB} & = & 0 \\
S_{AB}^{AB} & = & \beta\mbox{e}^{ip_2}d_{21}/g_{AB} =
 {\al \beta g_{A0}+g_{0B}e^{-ip_1-ip_2}   \over  \al g_{AB} e^{-ip_2} } \\
S_{BA}^{BA} & = & \beta^{-1}\mbox{e}^{-ip_1}g_{AB}/d_{12} =
{  \al g_{AB} e^{-ip_1} \over \al \beta g_{A0}+g_{0B}e^{-ip_1-ip_2} } \\
\label{S_caseII_end}
S_{BB}^{BB} & = & -\frac{b_{21}}{b_{12}}=
-{e^{-ip_1} \left(\beta- e^{-ip_2} \right) \over
e^{-ip_2} \left(\beta- e^{-ip_1} \right)} .
\eea

Notice that the corresponding $S$-matrix is diagonal. The unitarity
condition (\ref{unitary}) holds. The crossing-symmetry relation 
is generally  not satisfied.

\section{Dynamical algebra and Yang-Baxter equations}

The quadratic relations discussed above are sufficient to describe only
the sectors with one or two particles respectively. In order to study
the $n$-body problem within this approach one has to make sure that the 
expectation values expressed in terms of the quantities 
$\mbox{Tr } (D^Z_{p_1}D^{Z'}_{p_2}D^{Z''}_{p_3}\dots  E^{L})/Z_L$
automatically satisfy the original master equation, irrespective of the
contours of integration over the pseudo momenta $p_i$. This can only be
ensured by requiring associativity of the algebra defined by (\ref{algebra7})
which in turn implies conditions on the properties of the $\Sigma$-matrix
(\ref{sigma}) or the $S$-matrix respectively.

The matrix $\Sigma$ acts like a generalized permutation operator on the tensor
product of vector spaces defined by (\ref{ovec}). Hence associativity implies
that different orders of permutations must lead to the same final result.
This in not trivial since the vector components are non-commutative
objects. Let us define $\Sigma^{(1)}(p_1,p_2)=\Sigma(p_1,p_2) \otimes {\bf 1}$
with the $2\times 2$ identity matrix ${\bf 1}$ acting trivially on the third
subspace of a tensor vector ${\cal D}_{p_1} \otimes {\cal D}_{p_2} \otimes
{\cal D}_{p_3}$. Analogously we define $\Sigma^{(2)}$ as acting trivially on
the first subspace. Applying (\ref{algebra7}) in the order $1\leftrightarrow
2$,  $2\leftrightarrow 3$, $1\leftrightarrow 2$ yields
\be
{\cal D}_{p_1} \otimes {\cal D}_{p_2} \otimes {\cal D}_{p_3}
= \Sigma^{(1)}(p_1,p_2)\Sigma^{(2)}(p_1,p_3)\Sigma^{(1)}(p_2,p_3){\cal
D}_{p_3} \otimes {\cal D}_{p_2} \otimes {\cal D}_{p_1}
\ee
On the other hand, choosing the order of permutations as $2\leftrightarrow 3$,
$1\leftrightarrow 2$, $2\leftrightarrow 3$ one arrives at
\be
{\cal D}_{p_1} \otimes {\cal D}_{p_2} \otimes {\cal D}_{p_3}
=
\Sigma^{(2)}(p_2,p_3)\Sigma^{(1)}(p_1,p_3)\Sigma^{(2)}(p_1,p_2)
{\cal D}_{p_3} \otimes {\cal D}_{p_2} \otimes {\cal D}_{p_1} .
\ee

Associativity therefore implies
\bel{constraint3}
\Sigma^{(1)}(p_1,p_2)\Sigma^{(2)}(p_1,p_3)\Sigma^{(1)}(p_2,p_3) =
\Sigma^{(2)}(p_2,p_3)\Sigma^{(1)}(p_1,p_3)\Sigma^{(2)}(p_1,p_2).
\ee
This is set of 64 equations for the hopping rates which must be
satisfied for all $p_i$. (Extra solutions for special values are
discussed below). In terms of the elements of the $S$-matrix the
relations (\ref{constraint3}) read
\bel{YBE}
S_{ij}^{i''j''}(p_1,p_2) S_{i''k}^{i'k''}(p_1,p_3) S_{j''k''}^{j'k'}(p_2,p_3) =
S_{jk}^{j''k''}(p_2,p_3) S_{ik''}^{i''k'}(p_1,p_3) S_{i''j''}^{i'j'}(p_1,p_2) .
\ee
These are the Yang-Baxter equations with the usual Einstein convention of
summing over internal indices. In
another compact form they may be written
\bel{YBE2}
S_{12} S_{13} S_{23} = S_{23} S_{13} S_{12}
\ee
where $S_{ij}$ is the $S$-matrix acting on spaces $i,j$ as a function
of the pseudomomenta $p_i,p_j$. If these equations are satisfied
no extra constraints arise from consistency relations involving more
than three operators ${\cal D}$.

Since in case II the $S$-matrix is diagonal the YBE is satisfied 
automatically, i.e., there is no further constraint on the hopping rates
which would be required for integrability. In case I some more discussion
is necessary. The various conditions on the rates arising from the YBE
can be obtained analytically using the software packages $mathematica$
or $maple$.  
Due to the fact that there is a charge conservation 
( $ S_{ij}^{i'j'} =0$, unless $i+j=i'+j'$ ), only the equations (\ref{YBE})
with $i+j+k = i'+j'+k'$ are nonzero, which leaves 20 equations. Out of them
6 equations are satisfied trivially. The remaining  14 equations are pairwise
equivalent,  leaving only 7 independent 
relations. 

It is convenient to pick one of the those independent equations, derive  
the arising constraints on the rates (if any), use this constraint
in the remaining equations and then to iterate until all equations
are satisfied. The solutions that we have found can be classified according to the 
values of the hopping rates $g_{ZZ'}; \ Z,Z'=0,A,B$. 
 The crossing-symmetry relation 
(written as 
$S_{\al \beta}^{\al' \beta'}(p,q) =S_{\beta \bar{\al'}}^{\beta'\bar{\al} }(-p,-q)$,
 $\bar{A}=B, \bar{B}=A$
for the spectral-parameter dependent $S$ matrix  (\ref{S-matrix}) )
is in general  not satisfied. 
In that respect the algebra (\ref{scat})
is not exactly  the Zamolodchikov algebra \cite{Zamolodchikov}, but the 
one with some field-theoretical restrictions relaxed. 

\subsection{All hopping rates nonzero }
This group consist of a solution with
\bel{anisotropic_a}
(a)\  \al=\beta; \ \   g_{A0}=g_{B0}=g_{AB}=g; \ \ g_{0A}=g_{0B}=g_{BA}=h
\ee
and those obtained by relabeling the particles/holes,  $A \leftrightarrow B$, 
 $B \leftrightarrow 0$:
\bel{anisotropic_b}
(b)\  \al=\beta; \ \   g_{A0}=g_{B0}=g_{BA}=g; \ \ g_{0A}=g_{0B}=g_{AB}=h
\ee
\bel{anisotropic_c}
(c)\  \al=\beta h/g; \ \  
 g_{A0}=g_{0B}=g_{AB}=g; \ \ g_{0A}=g_{B0}=g_{BA}=h
\ee
Note that other reshuffling of labels will not result in new sets of rates, e.g. 
 relabeling  $A \leftrightarrow 0$  in (\ref{anisotropic_a}) gives again 
(\ref{anisotropic_a}). We have checked that all the cases
 (\ref{anisotropic_a}-\ref{anisotropic_c}) 
 lead to the  $S$-matrix of the same type given below. 
Therefore only the solution 
(\ref{anisotropic_a}) will be considered in detail.

Note that  constants $ \al,\beta$ are defined up to common factor, since  
one can redefine  $\al \mbox{e}^{ip} \rightarrow \mbox{e}^{ip}$.
Therefore in case (\ref{anisotropic_a}) one can consider $\al=\beta=1$ 
without losing generality.

The corresponding $S$-matrix 
\bel{S} 
 S( p_1, p_2) = 
\left(
\ba {cccc}
-{ K_{12}-(h+g)\mbox{e}^{-i p_1} \over  K_{12}-(h+g)\mbox{e}^{-i p_2} } & 0 &0 &0   \\
 0 &h {\mbox{e}^{-i p_1}-\mbox{e}^{-i p_2} \over K_{12}-(h+g)\mbox{e}^{-i p_2}} &
 -{K_{12}-g \mbox{e}^{-i p_2} - h \mbox{e}^{-i p_1} \over K_{12}-(h+g)\mbox{e}^{-i p_2}} &0   \\
 0 & -{K_{12}-g \mbox{e}^{-i p_1} - h \mbox{e}^{-i p_2} \over K_{12}-(h+g)\mbox{e}^{-i p_2}} &
 g { \mbox{e}^{-i p_1}-\mbox{e}^{-i p_2} \over K_{12}-(h+g)\mbox{e}^{-i p_2}}  &0   \\
 0 &0 &0 & -{K_{12}-(h+g)\mbox{e}^{-i p_1} \over K_{12}-(h+g)\mbox{e}^{-i p_2}} \\
\ea
\right)
\ee
$$K_{12}=h \mbox{e}^{-i p_1-i p_2} +g $$

after the transformation
\bel{parametrization}
\mbox{e}^{-i p}=  \mbox{e}^{\eta} \ \ 
{ \mbox{sinh}(\la) \over \mbox{sinh} (\la +\eta)  };
\ \ \  \mbox{e}^{\eta}=\sqrt{g \over h}
\ee
 is parametrized to a difference form 
$S(p_1,p_2) = S(\la_1 - \la_2)$. The precise form of $S$ as a
function of a spectral parameter $\la$ \cite{LiebWu}
$S( \la)$ is noteworthy  (sh $\equiv$ sinh):

\bel{S-matrix} 
{ \mbox{sh}(\la +\eta)} \ \ S( \la) = 
\left(
\ba {cccc}
\mbox{sh}(\la-\eta) & 0 &0 &0   \\
 0 &\mbox{e}^{-\eta } \mbox{sh} (\la) & -\mbox{e}^{\la} \mbox{sh} (\eta) &0   \\
 0 & -\mbox{e}^{-\la} \mbox{sh} (\eta) &\mbox{e}^{\eta } \mbox{sh} (\la) &0   \\
 0 &0 &0 &\mbox{sh}(\la-\eta)   \\
\ea
\right)
\ee
The anisotropy parameter plays a special role in the theory of 6-vertex
model \cite{LiebWu}
 \bel{6vDelta}
\Delta = {S^{11}_{11} S^{22}_{22} + S^{21}_{21} S^{12}_{12} - 
S^{12}_{21} S^{21}_{12} \over 
2 \sqrt{S^{11}_{11} S^{22}_{22} S^{21}_{21} S^{12}_{12 } } } = \mbox{cosh}(\eta)
\ee

If $h=g$, the transformation (\ref{parametrization}) is to be substituted
with $\mbox{e}^{-i p}= (\la+i/2)/  (\la-i/2)$, to retrieve the well-known rational 
solution of YBE $S(\la) \approx \la I +  i P$,  $P$ being a permutation operator.

Note however that for the general case $h \neq g$, the solution 
(\ref{S-matrix}) differs from the usual trigonometric one due to the 
spectral parameter dependence in the adiagonal elements. This dependence 
can be removed however by a similarity transformation, and in addition it
plays no role in the equations for the spectrum.
Note also that the sum of S-matrix elements along each column is the same for each 
column, so that $S$-matrix is stochastic.

Finally, note that the choice of rates (\ref{anisotropic_a}) 
was listed as an integrable case in \cite{Dahmen}.

\subsection{Some  hopping rates zero}
Note that Eq.(\ref{TA},\ref{TA1}) imply that both auxiliary constants
$\al, \beta$ are nonzero. In what follows, we require this to hold, 
 $\al \beta \neq 0$. 

1. $g_{AB}=g_{BA}=0, \ g_{A0}=g_{B0}, g_{0A}=g_{0B}, \al=\beta$.

Again as in the case (\ref{anisotropic_a}), one can choose $\al=1$.
The corresponding $S$-matrix is proportional to a permutation operator,
$S_{12} =f_{12} P$, defined as $P_{i' j'}^{i j} = \delta_{ij'} \delta_{i'j}$.
This is a tracer diffusion process. One can imagine it as usual 
exclusion process with particles $A$ and $B$ having identical dynamics,
but  different colors \cite{tracer_diffusion}.
 The proportionality coefficient is
\bel{f}
f_{12}=(f_{21})^{-1} = 
-{ g_{A0}- ( g_{0B}+  g_{A0}) \mbox{e}^{-i p_1} +  g_{0B} \mbox{e}^{-i p_1-i p_2}  
    \over 
g_{A0}- ( g_{0B}+  g_{A0}) \mbox{e}^{-i p_2} +  g_{0B} \mbox{e}^{-i p_1-i p_2} }
\ee

2. $g_{B0}=g_{A0}=g_{BA}=0, \ g_{0A}=g_{0B} \al/\beta,\ 
 g_{AB}=g_{0B} (\beta-\alpha)/\beta$.

Note that $g_{0B}=g_{0A}+g_{AB}$,  
and since the  constants $\al,\beta $ are arbitrary,  
 after relabeling holes and particles $A\leftrightarrow 0$
we obtain the set of constants considered already in {\it Case II}.
However since it is the different physical system, we shall list it independently 
here.  

The $S$-matrix is:

\bel{5v-matrix} 
- S(p_1,p_2) = \left(
\ba {cccc}
{1-\al \mbox{e}^{ip_2} \over 1-\al \mbox{e}^{ip_1}} & 0 &0 &0   \\
 0 & 0 &{1-\al \mbox{e}^{ip_2} \over 1-\al \mbox{e}^{ip_1}} &0   \\
 0 &{1-\al \mbox{e}^{ip_2} \over 1-\al \mbox{e}^{ip_1}} &
 { \al (\al-\beta) \over \beta } { \mbox{e}^{ip_2}- \mbox{e}^{ip_1} \over (  1-\al \mbox{e}^{ip_1})^2  } &0   \\
 0 &0 &0 & {1-\beta \mbox{e}^{ip_2} \over 1-\beta \mbox{e}^{ip_1}}\\
\ea
\right)
\ee
The $S$-matrix above can be viewed as the 5-vertex model.
The characteristic parameter $\Delta$ for this case is  \cite{DKim}:
\bel{Delta}
\Delta= { S_{AA}^{AA} S_{BB}^{BB} - S_{AB}^{BA} S_{BA}^{AB}
\over 
S_{BB}^{BB} S_{BA}^{BA} } = 
{ b (a-\mbox{e}^{-ip_2}) \over  a (b-\mbox{e}^{-ip_2})}
\ee
depends on the value of $p_2$. In the spirit of  
(\ref{tau}) the solution of the problem can be reduced to one of finding the 
eigen-values of the transfer-matrix with site-dependent weights. 
The essential property of the above $S(p_1,p_2) $-matrix is that it cannot be transformed
into a form containing only the difference of the spectral parameters, 
unlike the case (\ref{S-matrix}).
 The corresponding
Bethe equations are given in the next section. 
Alternatively, one can  proceed by relabeling
 particles and holes  $A\leftrightarrow 0$ and then using 
({\ref{BA_caseII}}). We assume that $\al \neq \beta$ in (\ref{5v-matrix})
since  $\al =\beta$ falls 
into  special tracer diffusion case (solution 1). 

\bigskip 

3. $g_{0A}=g_{0B}=g_{BA}=0, \ g_{B0}=g_{A0} \al/\beta,\  
 g_{AB}=g_{A0} (\beta-\alpha)/\beta$.
This model can be transformed into the previous one by renaming $A \leftrightarrow B$
and reversing the direction of particle motion  $g_{0A} \leftrightarrow g_{A0}$, etc..
The $S$-matrix is of type (\ref{5v-matrix} ).
Consequently the  equations for the spectrum are  analogous to (\ref{5vBA1}) 
and we shall not separately list them here. 

\bigskip 

The cases listed above exhaust the list of the nontrivial solutions of
the YBE for two species of particles.

\section{Spectral equations}

Because of particle number conservation eigenstates and their eigenvalues
can be classified according to number $N^Z$ of particles of each species that
move on the ring. We denote by $N=N^A+N^B$ the total number of particles.
The quantity $Q = N^A-N^B$ shall be referred to as charge.
\\[4mm]
$N=0$:\\[4mm]
This is the empty lattice. Since this is obviously an invariant state
under the stochastic dynamics. Hence the single eigenvalue 
\be
\epsilon=0 
\ee
of $H$ in this sector vanishes by construction.\\[4mm]
$N=1$:\\[4mm]
In order to obtain the relaxation spectrum from the eigenvalues
\be
\epsilon^Z_p =   g_{0Z} \gamma^{-1} \mbox{e}^{-ip} + g_{Z0} \gamma
\mbox{e}^{ip} -  g_{0Z} - g_{Z0} 
\ee
where $\gamma= \alpha,\beta$, $Z=A,B$ depending on the species of a
single-particle system on a finite lattice, one uses the Fourier ansatz and 
the solution  (\ref{solution}). This requires calculating ${\cal F}_Z(p) =
\mbox{Tr } ({\cal D}_Z(p) E^{L} Q)$ at time $t=0$.  Because of the cyclic
property of the trace the equation ${\cal F}_Z(p) = \gamma^L
\mbox{e}^{ipL} {\cal F}_Z(p)$ must be satisfied. This 
yields the spectral equation
\be
\gamma^L\mbox{e}^{ipL} = 1
\ee
and thus fixes the allowed values of the pseudomomenta $p=2\pi n/L + i
\ln{\gamma}$.
 We stress that this quantization of the pseudomomentum is {\em
not} a property of the matrices ${\cal D}_Q(p)$ themselves. It appears only as
a result of taking the trace in the one-particle sector. For expectation
values in higher sectors we shall obtain different quantization constraints.
We remark that for calculating only the spectrum the normalization factor
$Z_L$ is not required.\\[4mm] $N \geq 2$:\\[4mm]

In the case when there are $n$ particles of both types $A$ and $B$ in the system,
the averages are written in terms of quantities
\bel{AV}
{\cal F}^{Z_1,\ldots Z_N}(p_1,\ldots p_N) =
\mbox{Tr } ({\cal D}^{Z_1}(p_1) \ldots {\cal D}^{Z_N}(p_N)
 E^{L} Q)
\ee

Now we have to distinquish  between different integrable models.
\subsection* {Case II}

The easiest to handle is the {\it Case II}, where $S$-matrix
is diagonal.

Commuting the  ${\cal D}(p_k)$ through around the ``circle'' inside
the trace in (\ref{AV}),
and using the algebra (\ref{scat}), one gets, using the trace property:
\bel{BA_caseII}
\prod_{j \neq k}^{N} S^{Z_{k} Z_{j}}_{Z_{k} Z_{j}}(p_k,p_{j})
 \mbox{e}^{i p_k L}=1, \ \ k=1, \ldots N
\ee
with $S$-matrix elements written in 
(\ref{S_caseII_begin}-\ref{S_caseII_end}). $\al,\beta$ can be set both 
to $1$ by simple rescaling  of quasiimpulses of $A$ and $B$ particles.
It can be shown \cite{Alcaraz_private} that applying the
coordinate Bethe Ansatz directly to the process (\ref{constraint2_process})
yields the same 
result (\ref{BA_caseII}).

\subsection* {Case I, all nonzero hopping rates}

Commuting the  ${\cal D}(p_1)$ through around the ``circle'' inside
the trace in (\ref{AV}),
and using the algebra (\ref{scat}), 
one obtains:
\bel{tau_index}
S^{Z_1 Z_{2}}_{\al_{2} \gam{2}}(p_1,p_{2})
\prod_{j=3}^{N-1} S^{\al_{j-1} Z_{j}}_{\al_{j} \gam{j}}(p_1,p_{j})
 S^{\al_{N-1} Z_{N}}_{\gam_{1} \gam{N}}(p_1,p_{N})
 \mbox{e}^{i p_1 L}
{\cal F}^{\gam_1 \gam_2 \ldots \gam_N } ={\cal F}^{Z_1 Z_2 \ldots Z_N }
\ee
(summation over repeated indices is implied), 
that can be shortly rewritten in matrix form 
as 
\bel{tau}
-\mbox{Tr}_0 ( L_{01}(p_1,p_1) L_{02}(p_1,p_2) \ldots L_{0N}(p_1,p_N) )
{\cal F}_N = \mbox{e}^{-i p_k L } {\cal F}_N
\ee
using the property
$S^{\al \beta}_{\al' \beta'}(p,p)= -\de_{\al \beta'} \de_{\beta \al'} $ 
of the $S$-matrix (\ref{S}).
The matrix $L_{0k}(p_1,p_2)$ is a matrix (\ref{S}) acting nontrivially 
in $ su(2)_0 \otimes su(2)_k$, and acting as identity matrix in the other subspaces
from   
$su(2)_0 \prod_{j=1}^N \otimes su(2)_j $, and $\mbox{Tr}_0$ denotes 
trace over the $su(2)_0$.

Subsequent analysis of the above eigenvalue equations can be done
in the framework of standard coordinate \cite{Yang67} or algebraic nested 
\cite{Kulish} Bethe Ansatz, leading to following spectral equations
(we made the transformation (\ref{parametrization}) to the difference form): 

\be
\label{BA1}
\mbox{e}^{\eta L} \left(
{sh ( \la_k ) \over sh ( \la_k + \eta )}
 \right)^L = 
\left( -1 \right)^{N+1}
\prod_{n=1}^N 
{sh ( \la_k -\la_n -\eta ) \over  sh ( \la_n  -\la_k - \eta )}
 \prod_{\eps=1}^{N^B} 
{sh ( \la_\eps^{(1)} -\la_k -\eta ) \over  \mbox{e}^{\eta} 
sh  ( \la_\eps^{(1)} -\la_k )}
\ee
\be
\label {BA2}
 \mbox{e}^{-\eta N} \prod_{n=1}^N 
{sh ( \la_\nu^{(1)} -\la_n ) \over sh ( \la_\nu^{(1)} -\la_n - \eta) }
  = (-1)^{N_B+1} \prod_{\eps=1}^{N^B}
 {sh ( \la_\eps^{(1)}-\la_\nu^{(1)}-\eta) \over  
sh ( \la_\nu^{(1)}-\la_\eps^{(1)}-\eta)}
\ee
where $\mbox{e}^{\eta}= \sqrt{g/h},\  k=1,2\ldots N, \ \nu=1,2\ldots {N^B}, \ {N^B} \leq N$. The above Bethe Ansatz appeared without the proof in \cite{Dahmen}.

\subsection* {Case I, some zero hopping rates (tracer diffusion)}

Commuting the term depending on  $(p_k)$ the ``circle'' inside
the trace  in (\ref{AV}),
one has using the (\ref{AV},\ref{f}):
\bel{one_time}
{\cal F}^{Z_1,\ldots Z_N}(p_1,\ldots p_N) =
R_k(p_1, \ldots p_N)
{\cal F}^{Z_N Z_1\ldots Z_{N-1}}(p_1 p_2\ldots p_N) 
\ee
where
\bel{R_k}
R_k(p_1, \ldots p_N)=\prod_{j\neq k}^{N} f(p_k,p_{j}) \mbox{e}^{i p_k L}
\ee

The set of equations contained in
(\ref{one_time}) means that all $R_k$ are strictly the same,
 $R_1=R_2=\ldots R_N$. 
Note additionally that in the right-hand side of Eq.(\ref{one_time}),
the lower indexes are shifted one step to the the right. It gives the 
guideline for determining the spectral equation. E.g. if all $Z_i =A$,
shift of the sequence $ \{Z_i\}$ leaves it invariant, so  we will obtain
$R_k=1$, for any $k$. If  the sequence  $ \{Z_i\}$ is periodic with period n, 
recurrent use of  (\ref{one_time}) yields $R_k^n=1$. 
In the general case, when  $ \{Z_i\}$ is non-periodic, 
recurrent use of  (\ref{one_time}) $N$ times gives:
$$R_k(p_1, \ldots p_N)^N=1, \ \ R_1=R_2=\ldots R_N.$$

\subsection* {Case I, some zero hopping rates (5-vertex model)}

Proceeding analogously to (\ref{tau_index}), and denoting 
$a(p)=1-\al \mbox{e}^{ip},\ b(p)=1-\beta \mbox{e}^{ip}$
one  obtains:

\bel{5vBA1}
\left(
\al \mbox{e}^{ip_k}
\right)^L
 =
 \left( -1    \right)^{N+1}
\left( { \al \over \beta} -1   \right)^{N_B}
 \prod_{n=1}^N  
{a(p_k)  \over a(p_n)}
 \prod_{\eps=1}^{N^B}  
{  a(p_\eps^{(1)}) - a(p_k) \over
   a(p_\eps^{(1)})  a(p_k)   } 
\ee
\bel{5vBA2}
\left( { \beta  \over  \al  } \right)^{L}
\left( { \al \over \beta} -1   \right)^{N}
\prod_{n=1}^N  \ 
{  a(p_\nu^{(1)})-  a(p_n)\over
 a(p_\nu^{(1)}) a(p_n) }
=
 \left( -1    \right)^{N_B+1}
\prod_{\eps=1}^{N^B} 
 {  a(p_\eps^{(1)}) \over  a(p_\nu^{(1)})}
 {  b(p_\nu^{(1)}) \over  b(p_\eps^{(1)})}
\ee

where $ k=1,2\ldots N, \ \nu=1,2\ldots {N^B}, \ {N^B} \leq N$.

Alternatively, one can obtain the nested Bethe Anzatz in this and other
cases by 
 an additional (nested) Fourier transform of 
either ${\cal D}_A(p)$ or ${\cal D}_B(p)$, i.e. 
\bel{FT2}
{\cal D}_B(p)=  {\cal D}_A(p) \int \mbox{e}^{-i q p}  \Delta(q) d  q
\ee

We remark that the Zamolodchikov-type algebra not only leads to
the spectral equations (\ref{BA1})-(\ref{5vBA2}) but also 
implies functional relations for the matrix elements (\ref{AV}).
These functional relations are satisfied by Bethe wave functions
as in the spin $1/2$ case \cite{Schu98} and hence yield
expressions for the expectations as integrals over appropriately
chosen contour. Alternatively one could search for representations
\cite{Fring} and directly calculate the matrix elements. This
procedure requires further investigation.

\section*{Acknowledgments}

G.M.S. would like to thank R.B. Stinchcombe and F. Essler for useful
discussions and the Department of Physics, University of Oxford, where part of
this work was done for providing a stimulating environment. E.F. thanks the
Institut f\"ur  Festk\"orperforschung for kind hospitality. V.P.
acknowledges financial support from the Deutsche Forschungsgemeinschaft. 
We would like to thank F.C. Alcaraz for interesting discussions.

\bibliographystyle{unsrt}

\begin{thebibliography}{99}
\bibitem{xxx} For another choice of rates this model was studied in 
 \cite{DEHP}
\end{thebibliography}

\end{document}